\documentclass[aps,prc,twocolumn,a4paper,english,superscriptaddress,showpacs,showkeys]{revtex4-1}

\usepackage[ansinew]{inputenc}
\usepackage[T1]{fontenc}
\usepackage[english]{babel}
\usepackage[usenames,dvipsnames]{color}
\usepackage{graphicx}
\usepackage{amsmath}
\usepackage{amssymb}
\usepackage{hyperref}
\usepackage{multirow}

\newcommand{\code}[1]{#1}
\newcommand{\cxx}{\code{C++}}

\newcommand{\inclxx}{\code{INCL++}}
\newcommand{\inclf}{\code{INCL4.6}}

\newcommand{\incl}{\code{INCL}}

\newcommand{\isabel}{\code{Isabel}}
\newcommand{\bertini}{\code{Bertini}}
\newcommand{\cem}{\code{CEM03.03}}

\newcommand{\geminixx}{\code{GEMINI++}}
\newcommand{\abla}{\code{ABLA07}}

\newcommand{\ud}{\mathrm{d}}
\newcommand{\proton}{\textit{p}}

\newcommand{\lead}{$^\text{208}$Pb}
\newcommand{\calcium}{$^\text{40}$Ca}

\newcommand{\geant}{\code{Geant4}}

\newcommand{\etal}{\emph{et al.}}
\newcommand{\highlight}[1]{#1}
\newcommand{\highlightmissing}[1]{}

\ifpdf
  \graphicspath{{./figures/}}
\else
  \graphicspath{{./figures_eps/}}
\fi

\begin{document}
\title{Improving proton-induced one-nucleon removal in intranuclear
  cascade}

\author{Davide Mancusi}%
\email[corresponding author. E-mail address: ]{davide.mancusi@cea.fr}%
\affiliation{CEA, Centre de Saclay, Irfu/SPhN, 91191 Gif-sur-Yvette CEDEX, France}
\author{Alain Boudard}%
\affiliation{CEA, Centre de Saclay, Irfu/SPhN, 91191 Gif-sur-Yvette CEDEX, France}
\author{Jaume Carbonell}%
\affiliation{Institut de Physique Nucl\'eaire, Universit\'e Paris-Sud,
  IN2P3-CNRS, 91406 Orsay CEDEX, France}
\author{Joseph Cugnon}%
\affiliation{AGO department, University of Li\`{e}ge, all\'{e}e
  du 6 ao\^{u}t 17, b\^{a}t.~B5, B-4000 Li\`{e}ge 1, Belgium}
\author{{Jean-Christophe} David}%
\affiliation{CEA, Centre de Saclay, Irfu/SPhN, 91191 Gif-sur-Yvette CEDEX, France}
\author{Sylvie Leray}%
\affiliation{CEA, Centre de Saclay, Irfu/SPhN, 91191 Gif-sur-Yvette CEDEX, France}

\date{\today}

\begin{abstract}
  It is a well-established fact that intranuclear-cascade models generally
  overestimate the cross sections for one-proton removal from heavy, stable
  nuclei by a high-energy proton beam, but they yield reasonable predictions for
  one-neutron removal from the same nuclei and for one-nucleon removal from
  light targets. We use simple shell-model calculations to investigate the
  reasons of this deficiency. We find that a refined description of the neutron
  skin and of the energy density in the nuclear surface is crucial for the
  aforementioned observables, and that neither ingredient is sufficient if taken
  separately. As a by-product, the predictions for removal of several nucleons
  are also improved by the refined treatment.
\end{abstract}



\maketitle

\section{Introduction}
\label{sec:introduction}

Nuclear reactions between high-energy ($\gtrsim150$~MeV) nucleons or hadrons and
nuclei are usually described by means of hybrid models consisting of an
\emph{intranuclear-cascade} (INC) stage followed by a statistical de-excitation
stage \cite{filges-handbook}. In this framework, the projectile is assumed to
initiate an avalanche of binary collisions with the nucleons of the target,
which can lead to the emission of energetic particles. The nature of INC models
is essentially classical. It is typically assumed that nucleons are perfectly
localised in phase space and that they are bound by an average, constant
potential; moreover, it is assumed that subsequent elementary collisions are
independent.

Despite the simplicity of such reaction models, it has been proved that they are
able to describe a vast array of experimental observables with a very restricted
number of free parameters
\cite{leray-intercomparison,*intercomparison-website}. However, it was realized
some time ago that these models systematically fail to describe inclusive cross
sections for the removals of few nucleons \cite[for
example]{jacob-p2p,audirac-evaporation_cost}. This is especially surprising in
view of the fact that these observables are associated with peripheral reactions
and mostly involve collisions between quasi-free nucleons; one would therefore
expect intranuclear cascade to provide an accurate description of this
particular dynamics. This puzzling result has been known for many years now, but
no convincing explanation has ever been put forward.

Note that, in general, \highlight{the prediction of the inclusive
  one-nucleon-removal cross sections at high energy can reasonably be tackled
  only with a two-step dynamical/de-excitation model.}  One-nucleon removal, in
fact, results from events of a specific class: (1) few nucleon-nucleon
scatterings must take place, and (2) the residual excitation energies after
knockout must fall within a given window. If one of these conditions is not
verified, removal of several nucleons becomes the most likely outcome.  The
models that are usually applied to the study of knockout reactions
\cite{jackson-nuclear_reactions,froebrich-theory_nuclear_reactions} either do
not properly account for the probability of multiple collisions (condition 1) or
do not account for all the relevant residual states after knockout (condition
2).

One-nucleon removal, being associated with peripheral reactions, is certainly
sensitive to the details of the description of the nuclear surface, such as the
density profile. Arguably, the semi-classical initial conditions of INC might be
inadequate for this purpose. The aim of this paper is to investigate the
possibility to accommodate some genuine quantum-mechanical features of the
nuclear surface into INC, by appealing to simple shell-model calculations and by
casting their results in a form adaptable to INC. We will show to what extent
the predictions of a particular INC model \cite{mancusi-inclxx} can be thus
improved.

Section~\ref{sec:intranuclear-cascade} gives a brief description of the INC
framework, whose appropriateness for the problem at hand is specifically
discussed in Section~\ref{sec:appr-incde-excit}. The experimental data for
one-nucleon-removal cross sections are presented and interpreted in
Section~\ref{sec:one-nucleon-removal}. Our shell-model calculations are
described in Section~\ref{sec:refined-inc-model}, while the refined INC model is
introduced in Section~\ref{sec:refin-incl-nucl}. Results of the calculations
with the new model are presented in Section~\ref{sec:results}, before the
conclusions in Section~\ref{sec:conclusions}.

\section{Intranuclear cascade}\label{sec:intranuclear-cascade}

\subsection{Model description}\label{sec:model-description}

Intranuclear cascade \cite{serber-reactions} is a class of models that are
commonly used for the description of proton-induced reactions at high energy
($>150$~MeV). In this context it is assumed that the first stage of the reaction
can be described as an avalanche of independent binary collisions. The INC
scheme can be derived from the usual nuclear transport equations under suitable
approximations \cite{kadanoff-quantum_statistical_mechanics,bunakov-inc} and its
numerical solution can be efficiently tackled on today's computers. The INC
model is essentially classical, with the addition of a few suitable ingredients
that mimic genuine quantum-mechanical features of the initial conditions and of
the dynamics: for instance, target nucleons are endowed with Fermi motion,
realistic space densities are used, the output of binary collisions is random
and elementary nucleon-nucleon collisions are subject to Pauli blocking.

At the end of the intranuclear cascade, an excited remnant is left. This nucleus
typically relaxes by emitting low-energy particles or, when applicable, by
fissioning. The time-scale for the second stage is typically much longer than
that for the first one, which justifies the fact that de-excitation is not
described by INC but by a different class of models which rely on statistical
assumptions about the properties of the excited remnant. It is essential to
couple INC to a de-excitation model if one wishes to describe the production of
cold (i.e.\ observable) reaction residues.

INC approximates the exact dynamics of the nuclear reaction as a sequence of
binary collisions. However, the initial conditions of the reaction, which
typically amount to the ground state of the target nucleus and which are in
principle also determined by the exact nuclear dynamics, cannot be determined
within the INC approximation. It is therefore necessary to specify them as an
additional model ingredient.

In what follows, we make explicit reference to the Li\`ege Intranuclear Cascade
model \cite[\incl,][]{boudard-incl4.6} and the \abla{} statistical de-excitation
model \cite{kelic-abla07}. The \incl/\abla{} coupling is in general quite
successful at describing a vast number of observables in nucleon-induced
reactions at incident energies between $\sim60$ and $3000$~MeV
\cite{leray-intercomparison,*intercomparison-website}. For technical reasons,
the work described hereafter was performed with the latest \cxx\ version of the
\incl\ code \cite[\inclxx\ \code{v5.1.14},][]{mancusi-inclxx}. For the matter at
hand, \inclxx\ is essentially equivalent to the reference \inclf\ version
\cite{boudard-incl4.6}.

\begin{figure}
  \centering
  \includegraphics[width=\linewidth]{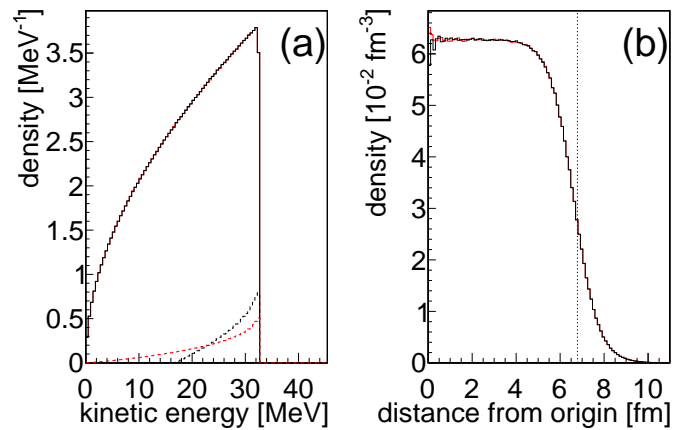}
  \caption{Proton kinetic-energy (a) and space (b) densities in \lead{} in the
    standard \incl{} initial conditions (solid black lines) and in
    the refined initial conditions with $f=0.5$ (solid red lines;
    see Section~\ref{sec:refin-incl-nucl} for the definition of $f$).
    The solid black and red lines are essentially on top of each
      other in both panels.  One-proton removal is dominated by impact
    parameters to the left of the vertical dotted line in panel (b). The dashed
    lines in panel (a) are the energy distributions of protons that are found to
    the right of the dotted vertical line of panel (b), for the
      standard (dashed black line) and refined (dashed red line) \incl{ initial
      conditions}.\label{fig:marginal-rp-correlation}}
\end{figure}

\begin{figure}
  \centering
  \includegraphics[width=\linewidth]{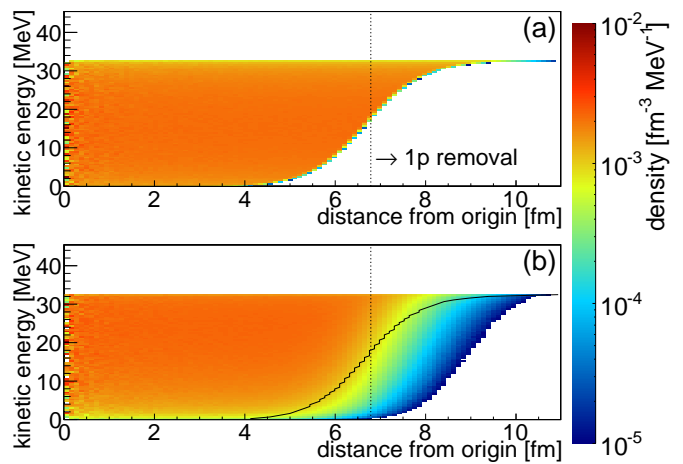}
  \caption{Space--kinetic-energy density of protons in \lead{} in the standard
    \incl\ initial conditions (a) and in the refined initial conditions with
    $f=0.5$ (b; see Section~\ref{sec:refin-incl-nucl} for the definition of
    $f$). The dotted vertical lines indicate the region of impact parameters
    which dominates one-proton removal. The contour of the colored shape in
    panel (a) represents the inverse of the function $R(T)$ and is reported as a
    solid black line in panel (b).\label{fig:rp-correlation}}
\end{figure}

The \incl{} model is peculiar in that it explicitly tracks the motion of all the
nucleons in the system, which are assumed to move freely in a square potential
well. The radius of the well is not the same for all nucleons, but it is rather
a function $R(T)$ of the nucleon kinetic energy (which is a conserved quantity
in absence of collisions). The initial nucleon momenta are uniformly distributed
in spheres of radii
\begin{align*}
  p_F(\text{proton})&={(2Z/A)}^{1/3}p_F\\
  p_F(\text{neutron})&={(2N/A)}^{1/3}p_F\text,
\end{align*}
with $p_F=270$~MeV${}/c$. The relation between kinetic energy and radius of the
potential well is such that the space density distribution is given by a fixed,
isospin-independent, suitable Woods-Saxon parametrisation \cite{boudard-incl}.

As an example, we discuss the phase-space density of protons in \lead, as
defined by the \incl{} initial conditions.
Figure~\ref{fig:marginal-rp-correlation} shows the proton kinetic-energy and
space distributions as solid black lines: the kinetic-energy distribution
represents a uniform sphere in momentum space, while the space distribution is
the classic Woods-Saxon distribution. Figure~\ref{fig:rp-correlation}(a)
illustrates how the kinetic energy correlates with the radius of the potential
well where the protons move. For a given kinetic energy, the density is constant
up to a certain distance from the center, which is the radius of the potential
well $R(T)$; beyond this radius, the density vanishes. The radius of the well
increases as the kinetic energy increases and reaches the radius of the
calculation sphere (inside which the simulation takes place; here about $11$~fm)
as the kinetic energy tends to the Fermi energy. The distributions shown in
Fig.~\ref{fig:marginal-rp-correlation} are simply projections of
Figure~\ref{fig:rp-correlation}(a) on each of the axes (up to multiplication by
appropriate Jacobians).

In substance, the motion of nucleons in the \incl{} nucleus is such that the
closer they are to the Fermi energy, the farther out they move in space; this
trait is inspired by the properties of classical particle motion in a potential
well. Therefore, the nuclear surface of the nucleus is predominantly populated
by nucleons whose energy is close to the Fermi energy.

\begin{figure*}
  \centering
  \begin{minipage}{0.49\linewidth}
    \includegraphics[width=\linewidth]{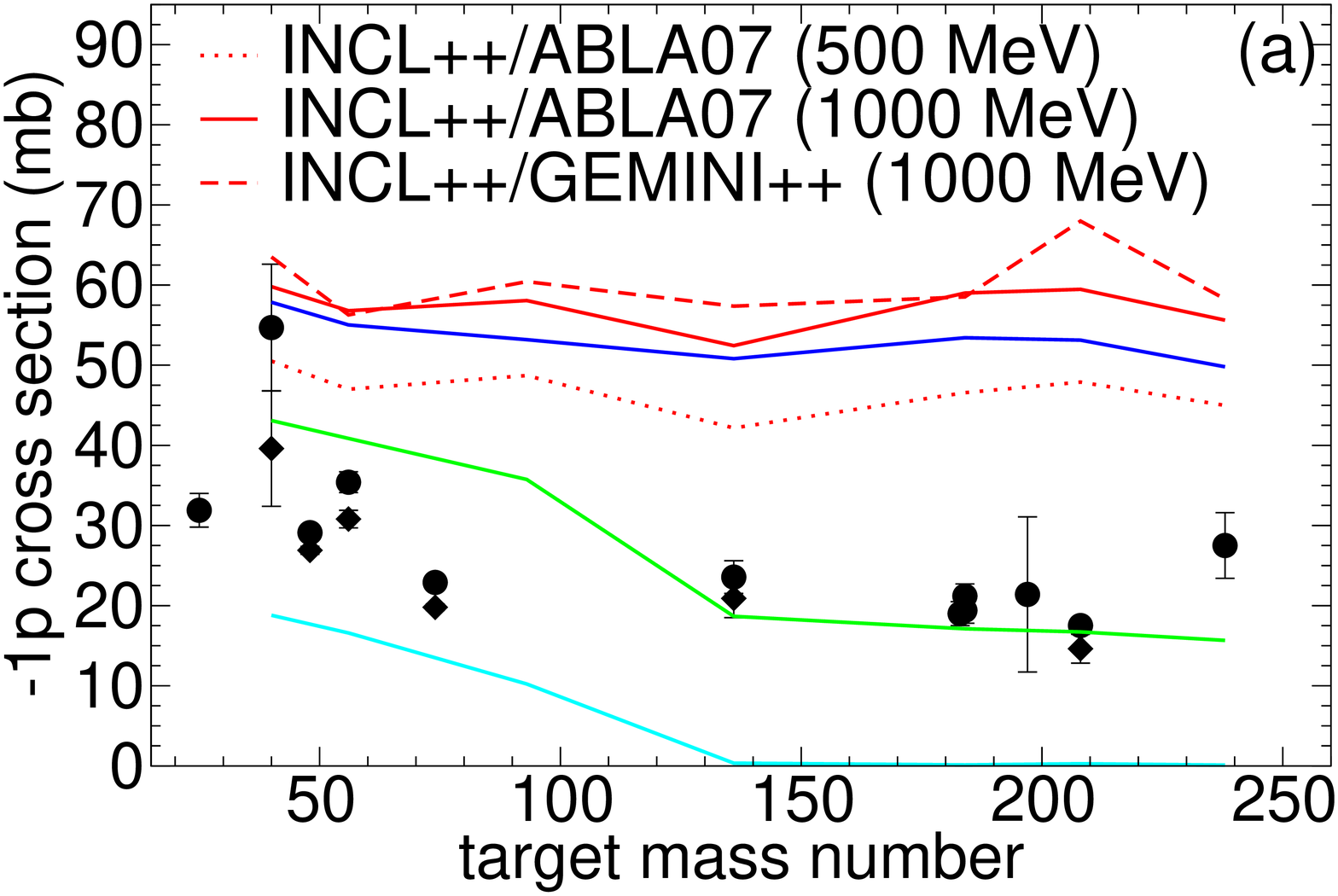}
  \end{minipage}\hfill%
  \begin{minipage}{0.49\linewidth}
    \includegraphics[width=\linewidth]{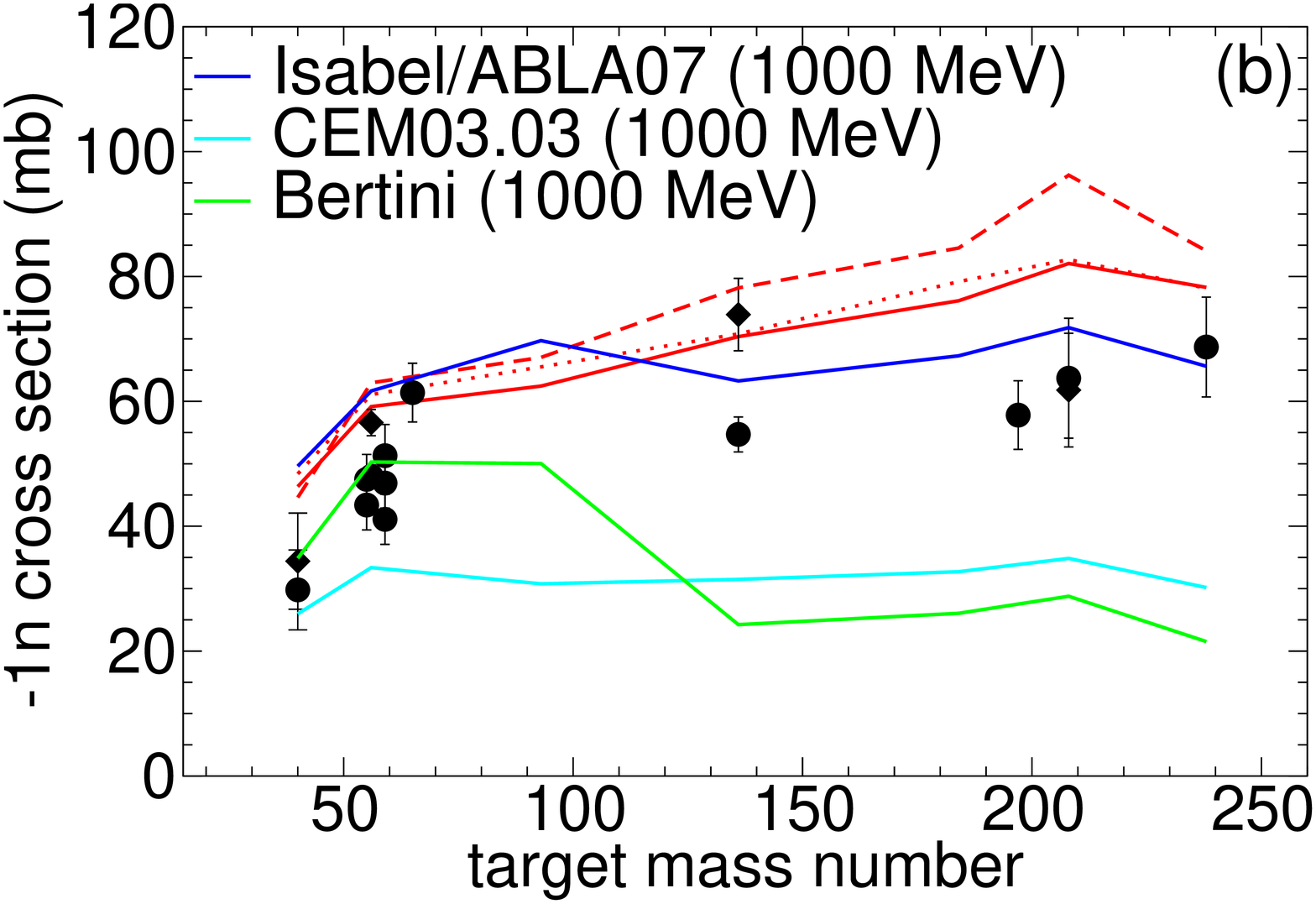}
  \end{minipage}
  \caption{Experimental data for one-proton- (a) and one-neutron-removal cross
    sections (b) in proton-nucleus reactions above 500~MeV incident energy, as a
    function of the target mass. Diamonds refer to experimental beam energies
    between 500 and 750~MeV, while circles represent energies above 750~MeV. The
    solid lines represent calculations with \incl/\abla{} (red), \isabel/\abla\
    (blue), \cem{} (cyan) and \geant's \bertini{} model (green) at 1000~MeV. The
    dashed red lines represent \incl/\geminixx\ calculations at 1000~MeV. The
    dotted red lines represent \incl/\abla{} calculations at
    500~MeV. Experimental data taken from
    Refs.~\citenum{chen-ca40,villagrasa-fe,rejmund-gold,giot-xe_500MeV,napolitani-xe_xsec,audouin-lead,enqvist-lead,taieb-u,titarenko-indc,michel-nuclide,jacob-p2p,reeder-mg25_p2p}.\label{fig:world-data}}
\end{figure*}

\subsection{Appropriateness of INC/de-excitation}\label{sec:appr-incde-excit}

As we mentioned in the Introduction, dynamical/de-excitation models are the only
ones that can reasonably attempt an inclusive description of one-nucleon removal
at high energy. \highlight{The dynamical stage may be described by INC (as in
  this paper) or by other models, such as the Boltzmann-Uehling-Uhlenbeck (BUU)
  or Vlasov-Uehling-Uhlenbeck (VUU) approaches \cite{remler-BUU}, or models from
  the family of quantum molecular dynamics (QMD)
  \cite{sorge-RQMD,aichelin-QMD,niita-jqmd}. These models are all ``INC-like''
  insofar as they superimpose a cascade of binary collisions on some kind of
  particle dynamics. For the sake of simplicity, in what follows we will always
  refer to INC models, but most of our analysis can be generalized to other
  classes of dynamical models.}

\highlight{Other kinds of nuclear-reaction models, such as} the distorted-wave
Born approximation (DWBA) or distorted-wave impulse approximation (DWIA)
\cite{jackson-nuclear_reactions,froebrich-theory_nuclear_reactions}, are not
expected to be applicable to the description of one-nucleon removal, for two
main reasons.

First, one-nucleon removal must be dominated by events with
few nucleon-nucleon scatterings, perhaps only one. This ensues from the fact
that the average energy transfer in nucleon-nucleon scattering at high energy is
large; therefore, multiple collisions are liable to lead to many-nucleon
removal. The probability for multiple collisions must therefore be correctly
evaluated. In DWBA and DWIA, rescattering is modeled as absorption owing to the
imaginary part of an optical potential, which seems far-fetched at the energies
we are concerned with. Analyses of proton spectra from the continuum have indeed
shown that, even in the 200--400~MeV incident-energy range, it is necessary to
go beyond DWIA insofar as the description of rescattering is concerned \cite[see
e.g.][]{pilcher-pp_c12,foertsch-pp_au197,cowley-pp_ca40}. Finally, the
evaluation of rescattering at higher energies is complicated by the possible
production of pions in the knockout collision.

Second, the one-nucleon-removal cross sections are fed by knockout reactions
leading to \emph{all} residue states below the particle-separation energy. 
In addition, the cross section can also receive contributions from the
continuum, i.e. from excitation energies above the particle-separation
energy. This is especially true for neutron removal, which can proceed through
the formation of an equilibrated system. The decay of the latter (and the
competition between the various channels) is accounted for in INC/de-excitation,
but not in DWBA/DWIA.

In any case, the INC/de-excitation \highlight{(dynamical/de-excitation)}
approach is the only one which can tackle with the same (combined) model all
residue formation channels, from one-nucleon removal channels to deep spallation
channels where a substantial fraction of the target nucleons are removed.

Admittedly, INC suffers for other limitations, the most important being the
semi-classical nature of the initial conditions.  This is reflected,
for instance, in the lack of a discrete level structure. The INC predictions are
typically smooth functions of the reaction parameters (charge, mass,
energy\ldots).  However, it is possible to allow for certain genuine
quantum-mechanical aspects in an effective manner. This is what we propose to
illustrate in Sections~\ref{sec:refined-inc-model} and
\ref{sec:refin-incl-nucl}.

\section{One-nucleon-removal}
\label{sec:one-nucleon-removal}

\subsection{Cross sections}
\label{sec:cross-sections}

Figure~\ref{fig:world-data} shows the experimental data for one-nucleon removal
in proton-induced reactions at energies $\gtrsim500$~MeV, as a function of the
target mass (all targets are close to the stability valley). Calculations with
\incl/\abla{} at 500 and 1000~MeV are shown for comparison. It is clear that the
model predictions are in the right ballpark for neutron removal, but they
overestimate the proton-removal data by a factor that can be as large as 3--4
for heavy nuclei. Note that the experimental data are globally consistent, even
though they have been collected in inverse-kinematics experiments
\cite{chen-ca40,villagrasa-fe,rejmund-gold,giot-xe_500MeV,napolitani-xe_xsec,audouin-lead,enqvist-lead,taieb-u}
or by off-line gamma spectroscopy
\cite{titarenko-indc,michel-nuclide,jacob-p2p,reeder-mg25_p2p}.

The role played by de-excitation can be clarified by comparing calculations with
the same INC but different de-excitation models. Therefore, we performed
calculations with \incl{} coupled with the \geminixx\ model
\cite{charity-gemini++,*mancusi-gemini++_fission}. The resulting cross sections
(shown in Fig.~\ref{fig:world-data} as dashed red lines) are within $20\%$ of
the \incl/\abla{} values and indicate that the influence of de-excitation is
rather mild.
Therefore, it seems unlikely that de-excitation can be held
  responsible for the gross overestimation of proton-removal cross sections.

We also show in Fig.~\ref{fig:world-data} the results of calculations that we
have performed with other INC models. The results of Isabel
\cite{yariv-isabel1}, in coupling with \abla, are qualitatively similar to the
\incl{} results and exhibit the same defect for proton removal. \geant's
Bertini-like cascade \cite{heikkinen-bertini_g4,*kelsey-bertini_preparation} is
quite different inasmuch as it yields rather correct predictions for proton
removal, but it badly underestimates neutron removal. Finally, \cem\
\cite{mashnik-cem} yields very low proton-removal cross sections (of the order
of $0.1$~mb for $A>130$) and underestimates neutron removal by roughly a factor
of two.

The calculations presented in Fig.~\ref{fig:world-data} globally demonstrate
that INC models have difficulty in correctly predicting the inclusive
one-nucleon-removal cross sections. \emph{No model} is able to describe proton-
and neutron-removal cross sections on all targets. This is rather surprising on
two counts. First, one-nucleon-removal cross sections are among the largest
isotopic cross sections, they are only modestly influenced by de-excitation and
they vary slowly with the target mass and the projectile energy; thus, they
represent an excellent test bench for INC models, but they seem to have
attracted little attention so far. Second, one-nucleon removal reactions are
typically dominated by very peripheral impact parameters, which probe regions of
the nucleus with large mean free path (low density); in addition, the collision
partners are somewhat localized in the nuclear surface, i.e.\ they are loosely
bound. One would expect the INC approximation to be fully justified under these
conditions. The failure illustrated by Fig.~\ref{fig:world-data} suggests that
INC models might be affected by a fundamental defect.

There are other remarkable features of the INC failure. One might expect even
more conspicuous mispredictions for the removal of a larger number of nucleons,
but one instead finds that the models can generally reproduce most of the
isotopic distributions rather well (see
the isotopic distributions in Sec.~\ref{sec:results} and
Ref.~\citenum{leray-intercomparison,*intercomparison-website}). This should be
understood as a consequence of the larger excitation energies associated with
the emission of several nucleons. Since large excitation energies can be
realized in numerous ways, some averaging takes place and the predictions become
less sensitive to the details of the initial conditions. At the same time, it
should be stressed that discrepancies do seem to increase for the removal of
e.g.\ several protons from stable nuclei \cite{audirac-evaporation_cost}. This
is consistent, inasmuch as the constraints on the excitation energy in that case
are even stricter than for one-proton removal, as evidenced by the smallness of
the associated cross sections.

Note that the experimental cross sections do not seem very sensitive to the
reaction parameters, such as the beam energy and the target species. This
suggests that the details of the level structure of the individual nuclides
involved do not play an important role. Therefore, it might be possible to amend
INC and describe these observables, but it is probably necessary to go beyond
the naive semi-classical model of the nuclear surface. We will do so in
Sections~\ref{sec:refined-inc-model} and \ref{sec:refin-incl-nucl}, but we first
need to clarify the mechanism that leads to proton and neutron removal within
the INC framework.

\subsection{Removal mechanism}\label{sec:removal-mechanism}

Let us first concentrate our attention on proton removal. The analysis of the
INC/de-excitation calculations indicates that proton removal is dominated (about
90\% of the cross section) by events with only one proton-proton collision. The
two protons leave the nucleus, which however retains some excitation energy. If
only one collision takes place, the excitation energy is simply given by the
depth of the proton hole, i.e.\ the difference between the Fermi energy and the
initial energy of the ejected proton. In any case, the excitation energy
remaining at the end of cascade is evacuated during the de-excitation phase.

Note that, for most $\beta$-stable, non-fissile nuclei, particle emission at low
excitation energy is largely dominated by neutron evaporation (for the sake of
illustration we neglect light nuclei, for which proton and $\alpha$ evaporation
can become competitive against neutron evaporation). If the excitation energy is
lower than the neutron separation energy $S_n$, no particle can be evaporated
and the energy will be evacuated as gamma rays. This is also true at energies
slightly larger than $S_n$, as long as gamma-ray emission outcompetes neutron
evaporation; thus, the effective neutron-evaporation threshold $S^*_n$ is
slightly larger than $S_n$. Therefore, the proton-removal channel is populated
\emph{if and only if} exactly one proton was ejected during INC and the
excitation energies at the end of cascade lies below $S^*_n$. If the excitation
energy allows for neutron evaporation, the final residue will be lighter (target
minus one proton minus $x$ neutrons).

The observations above highlight two important aspects. First, the
proton-removal cross section is extremely sensitive to the excitation energy
left in the nucleus after the ejection of a proton during INC. More precisely,
the cross section is determined by the probability that the ejection of a proton
during INC deposits an excitation energy smaller than $S^*_n$. Second, there is
a subtle difference between proton and neutron removal. Neutron removal can be
realized in two ways: either as a neutron ejection during INC followed by no
evaporation (this is analogous to the proton-removal mechanism), or as no
neutron ejection during INC followed by evaporation of one neutron. In the
latter scenario it is of course required that the incoming proton undergoes at
least one binary collision and that it succeeds in escaping from the target;
some conditions on the excitation energy also apply.

In either case, the fate of the de-excitation stage is essentially determined by
the excitation energy at the end of INC and by the neutron separation energies
in the region of the nuclide chart around the target. In this sense, our results
are essentially independent of the choice of the de-excitation model, as far as
all of them employ very similar separation energies for stable nuclei. The
second-order dependence on the de-excitation model (see the \abla/\geminixx\
difference) can be ascribed to differences in the neutron-gamma competition,
i.e.\ by slightly different values of the effective neutron-evaporation
thresholds $S^*_n$.

It is clear then that the smallness of the excitation energy at the
end of INC, especially in the proton-removal case, is the crucial
element that determines one-nucleon-removal cross sections. Comparison with the
experimental data (Fig.~\ref{fig:world-data}) seems to suggest that \incl\
largely underestimates the excitation energy associated with the ejection of a
proton. Similar remarks have been made about ``cold fragmentation'' in
peripheral nucleus-nucleus reactions \cite{benlliure-cofra}. It was found that
the excitation energy predicted by the abrasion model needs to be multiplied by
roughly a factor of two to explain the cross sections for the removal of one or
more protons. Given the simple nature of the abrasion model, however, the
generality of this conclusion is unclear.

Some important remarks are due at this point. Although \incl{} and \isabel{} are
in quantitative disagreement with the experimental data
(Fig.~\ref{fig:world-data}), they correctly capture the overall dependence on
the target mass. \bertini{} and \cem, on the other hand, yield trends that are
sensibly different from the experimental ones. This seems to point to the
existence of two classes of models and might be correlated with the presence of
an intermediate pre-equilibrium stage in \bertini{} and \cem. Because
one-nucleon removal is essentially a surface process, it is rather sensitive to
the geometrical arrangement of the first nucleon-nucleon collision. Treating one
of the collision partners as a pre-equilibrium exciton amounts to discarding all
information about its localization in configuration space. This probably entails
lower emission probabilities and higher excitation energies compared to a full
INC treatment, and might explain \cem's low proton-removal cross sections.

For completeness' sake, one should also remark that the two classes of models
may also be characterized by the nature of the INC stage. On the one hand,
\bertini{} and \cem{} are ``space-like'' INC models, i.e.\ they sequentially
track cascading nucleons until either they escape or their energy falls below a
given level. On the other hand, \incl{} and \isabel{} are ``time-like'' INC
models, i.e.\ they simultaneously track all cascading nucleons according to a
global clock. Nevertheless, one would not expect important differences between
these two approaches for reactions involving such a small number of cascading
nucleons. It is doubtful to us that this element can explain the contrasting
cross-section trends.

Finally, note that there are other ingredients which could in principle affect
the one-nucleon removal cross sections, such as the cross section of the first
nucleon-nucleon collision, its kinematics, the parameters of the nuclear-density
function (radius and diffuseness), the value of the Fermi momentum, the depth of
the nucleon potential well, the height of the Coulomb barrier assumed in INC and
in de-excitation and the separation energies during the INC stage. We have
verified that reasonable changes in these ingredients either have a negligible
effect on the calculated one-nucleon-removal cross sections or degrade the
agreement for neutron removal.

It is worthwhile at this point to summarize the statement of the problem and our
motivation. One has a rather successful semi-classical model for spallation
reactions, which can describe with the same ingredients channels resulting from
the ejection of few particles, as well as those corresponding to the emission of
a substantial part of target nucleons. This model seems to clearly fail on a few
channels, basically the one-nucleon removal channels. We gave arguments
indicating that this is due to the fact that these channels correspond to a
single-scattering mechanism leaving the target with a small excitation. We
pointed out that the semi-classical nature of the model is too crude to give
proper control of this excitation energy.  In the following, we will illustrate
a method to cure the deficiencies of INC on this point.

\section{Shell-model study of the nuclear surface}
\label{sec:refined-inc-model}

We mentioned at the end of Section~\ref{sec:model-description} that the nuclear
surface in the \incl{} initial conditions is predominantly populated
by nucleons whose energy is close to the Fermi energy. The ejection of one such
nucleon during INC results in little excitation energy for the cascade
remnant. The considerations in the previous section cast some suspicion upon
this aspect.

In the quantum-mechanical square-well problem, the density outside the well does
not vanish, even for states close to the bottom of the well. This means that
there is a non-zero probability to find deeply bound particles outside the well,
and eject them. This genuine quantum phenomenon is missing in the naive INC
nuclear picture, as illustrated by Fig.~\ref{fig:rp-correlation}(a). However, a
word of caution is due.  In a purely quantum-mechanical treatment, the surface
diffuseness is at least partly due to the penetration of the nucleon
wavefunctions into the classically forbidden region; in spite of this, the INC
initial conditions typically \emph{do} account for a realistic diffuseness of
the nuclear surface (e.g.\ the space density of the \incl{} model is a realistic
Woods-Saxon distribution), although this is entirely enforced by a
\emph{classical} correlation between the particle position and energy. The
failure of the INC initial conditions is therefore more subtle. It does not
concern the presence of the tail of the spatial density but rather its energy
density.

Another detail that is usually neglected in the INC picture is the presence of
neutron (or proton) skins in certain nuclei. It is for instance rather well
ascertained that \lead{} exhibits a neutron skin thickness (defined as the
difference of the neutron and proton root-mean-square radii) of about $0.2$~fm
\cite{zenihiro-neutron_skin}.  For grazing collisions, this means that the local
neutron density is several times larger than the proton density, leading to an
enhanced probability for collisions on neutrons.

\begin{figure}
  \includegraphics[width=\linewidth]{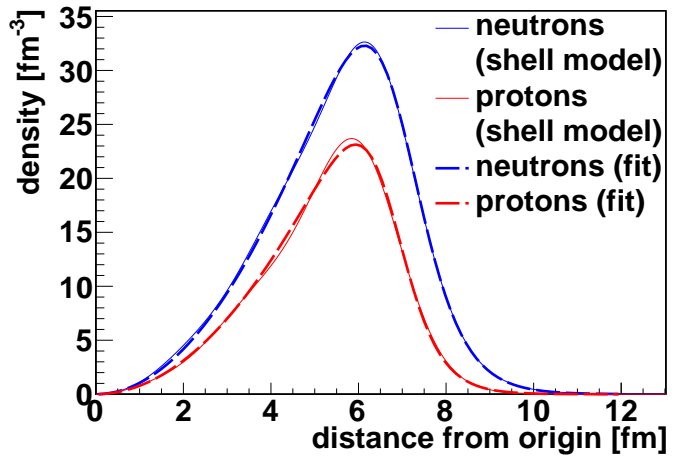}
  \caption{Proton (red) and neutron (blue) densities for \lead.  The thin solid
    lines represent the result of the shell-model calculation, while the thick
    dashed lines are Woods-Saxon fits. The fit parameters are given in
    Table~\ref{tab:density-fit}.  All curves include the Jacobian factor $4\pi
    r^2$. \label{fig:density}}
\end{figure}

We have estimated the magnitude of both the effects above with a simple
shell-model calculation. We assumed a central Woods-Saxon nuclear potential with
a spin-orbit term and a Coulomb term for the protons \cite{blomqvist-shell}. We
numerically solved the radial part of the Schr\"{o}dinger equation and
determined the radial eigenfunctions $R^i_{nj}(r)$ and the eigenvalues
$E^i_{nj}$ of the bound states (here $i=p,n$).  The single-particle energies
$E^i_{nj}$ in $^{208}$Pb correctly reproduce the energies of the lowest-lying
particle-hole states in $^{207,209}$Pb and $^{207}$Tl, $^{209}$Bi.

\subsection{Space densities}
\label{sec:space-densities}

In order to keep the notation simple, we drop the $i$ superscript from our
formulas and we consistently refer to protons (the formulas for neutrons can be
straightforwardly recovered). We assume that the shells are filled from the
bottom of the well up to the Fermi level. The latter may be partly empty if the
nucleus is not proton-magic. The occupation numbers $g_{nj}$ are given by
\[
g_{nj}=
\begin{cases}
  2j+1 & \text{if }E_{nj}<E_\text{F}\\
  Z-\sum_{E_{nj}<E_\text{F}}g_{nj} & \text{if }E_{nj}=E_F\\
  0 & \text{if }E_{nj}>E_F
\end{cases}\text.
\]
Here $E_\text{F}$ denotes the Fermi energy. We construct the radial density
profiles
\[
\rho_{nj}(r)=4\pi r^2|R_{nj}(r)|^2
\]
and the densities
\[
\rho(r)=\sum_{nj}g_{nj}\rho_{nj}(r)\text.
\]
The resulting proton and neutron densities are shown in
Fig.~\ref{fig:density}. At each position we can also construct the presence
probabilities
\begin{equation}
  p_{nj}(r)=g_{nj}\rho_{nj}(r)\left/\rho(r)\right.\text.
  \label{eq:presence}
\end{equation}

\begin{table}
  \caption{Optimal parameters for Woods-Saxon densities fitting the results of
    shell-model calculations. The ``skin/halo'' values are differences of the
    neutron and proton parameters. All values are in
    fm.\label{tab:density-fit}}
  \centering
  \begin{ruledtabular}
    \begin{tabular}{cc|cc|c}
      & & \multirow{2}{*}{neutrons} & \multirow{2}{*}{protons} & skin/\\
      & & & & halo\\
      \hline
      \multirow{2}{*}{\calcium} & $R_0$ & $3.57$ & $3.64$ & $-0.08$\\
      & $a$   & $0.49$ & $0.51$ & $-0.02$\\
      \hline
      \multirow{2}{*}{\lead} & $R_0$ & $6.98$ & $6.71$ & $0.26$\\
      & $a$   & $0.55$ & $0.46$ & $0.09$
    \end{tabular}
  \end{ruledtabular}
\end{table}

We would like to use the shell-model proton and neutron densities as inputs for
our INC calculation; however, the particle densities in \incl{} cannot be given
by an arbitrary function, so we must somehow adapt the shell-model densities. We
choose to fit them with Woods-Saxon distributions (shown in
Fig.~\ref{fig:density} as dashed lines). The best-fit parameters are indicated
in Table~\ref{tab:density-fit} and show that the shell-model densities for
\lead{} exhibit a neutron skin, although its thickness is slightly larger than
the experimentally accepted value; this is a well-known defect of mean-field
calculations \cite{friedman-exotic_atoms}. The calculations for \calcium\
instead yield a thin proton skin.

We thus decouple the \incl{} parameters describing the neutron space density
from those describing the proton space density. We choose not to modify the
proton densities (because they are already given by fits to the experimental
charge radii), but we adjust the neutron parameters by the amounts indicated in
the last column of Table~\ref{tab:density-fit}.

\subsection{Energy density of the nuclear surface}
\label{sec:energy-cont-nucl}

We have explained in the previous section that the outcome of single-collision
cascades is sensitive to the energy of the ejected nucleon. For a given
position, the shell model provides a decomposition of the local density in terms
of the various shells, Eq.~\eqref{eq:presence}.

In order to estimate the energy density of the surface, we assume that the
probability that a collision ejects a nucleon from a given shell is proportional
to the local density of the shell orbital. Furthermore, we neglect rearrangement
of the other nucleons in the Fermi sea after the collision; this amounts to
assuming that the excitation energy of the hole is simply given by the depth of
the hole:
\begin{equation}
  E^*_{nj}=E_\text{F}-E_{nj},\label{eq:estar_nj}
\end{equation}
where $nj$ are the quantum numbers of the hole. Putting all the pieces together,
we assume that a collision at position $r$ creates a hole of excitation energy
$E^*_{nj}$ [Eq.~\eqref{eq:estar_nj}] with probability $p_{nj}(r)$
[Eq.~\eqref{eq:presence}].


\begin{figure}
  \centering
  \includegraphics[width=\linewidth]{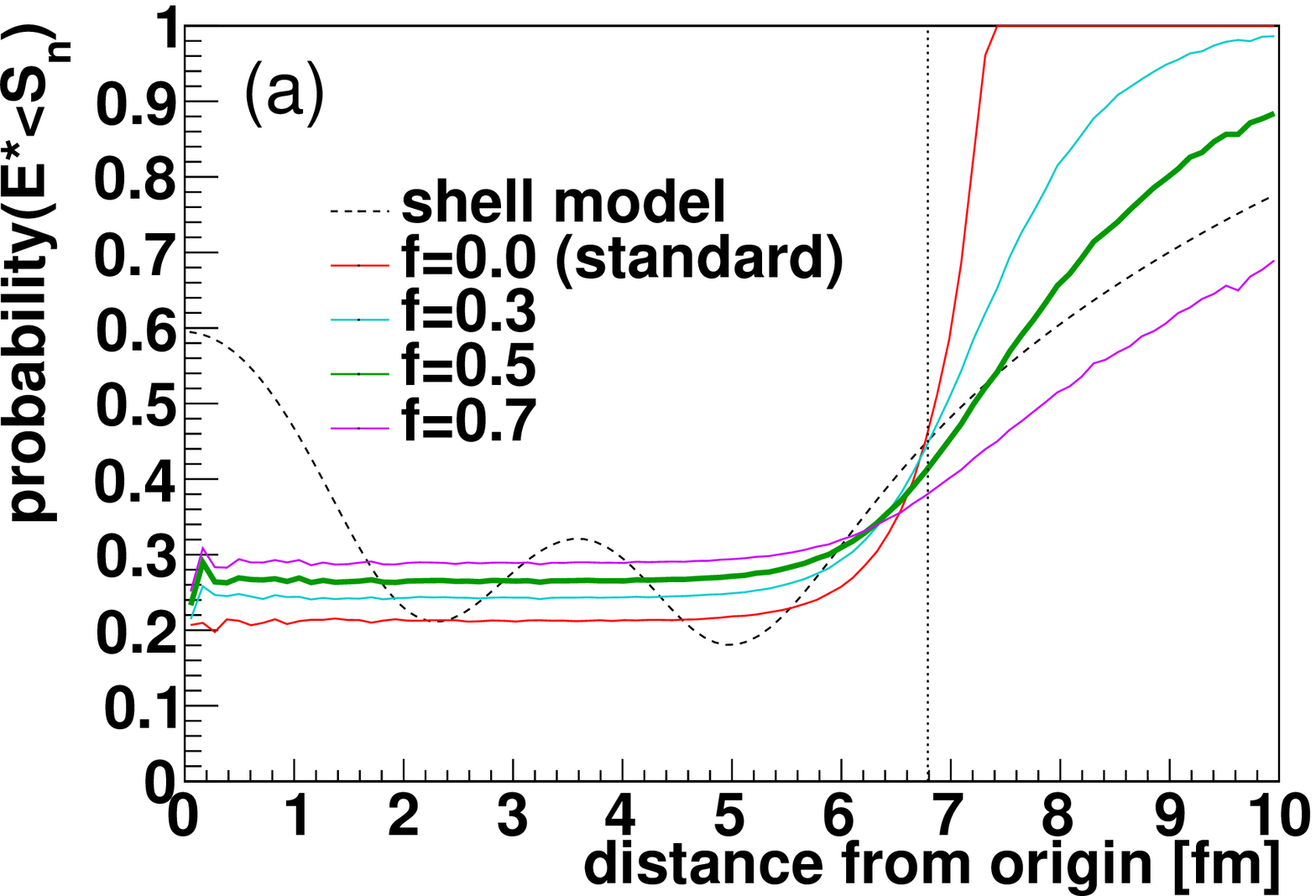}
  \includegraphics[width=\linewidth]{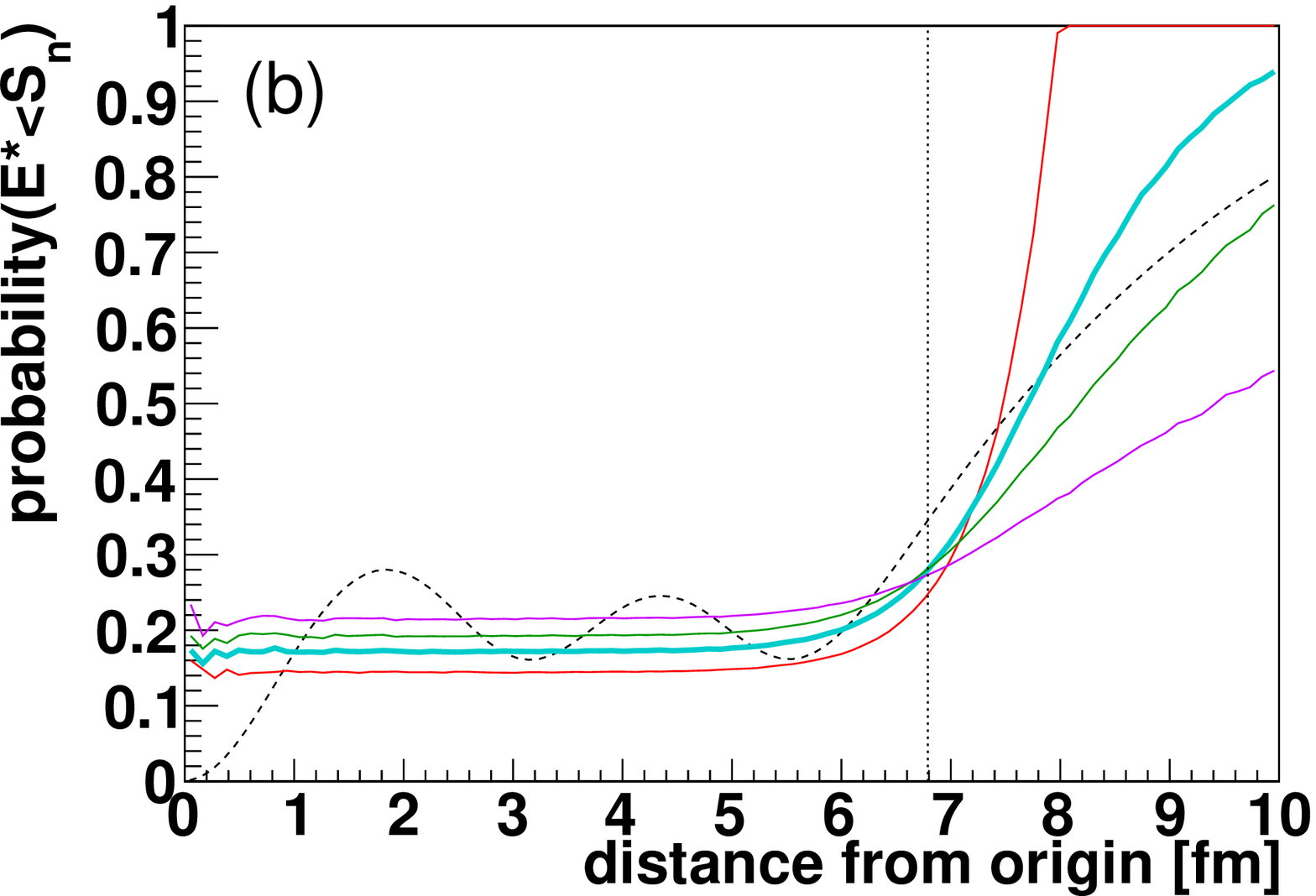}
  \caption{Probability that a proton (a) or neutron (b) hole in $^{208}$Pb
    results in an excitation energy smaller than the shell-model neutron
    separation energy, as a function of the distance of the hole from the center
    of the nucleus. The dashed line denotes the result of the shell-model
    calculation. Solid lines represent the \incl{} initial
      conditions for different values of the fuzziness parameter $f$ (defined
    in Section~\ref{sec:refin-incl-nucl}). The standard condition corresponds to
    $f=0$. One-proton removal is dominated by impact parameters to the left of
    the vertical dotted lines (see text).\label{fig:threshold}}
\end{figure}

We can characterize the properties of the nuclear surface by studying the
probability that the excitation energy associated with the hole does not exceed
the neutron separation energy, which reads
\[
P_{E^*<S_n}(r) = \sum_{nj} p_{nj}(r) \Theta(S_n-E^*_{nj})\text,
\]
where $\Theta$ is the Heaviside function.  From our discussion in
Section~\ref{sec:removal-mechanism} is should be clear that this quantity has a
very important bearing on one-nucleon-removal cross sections.

The probabilities for the standard \incl{} initial conditions are
plotted in Fig.~\ref{fig:threshold} as red solid lines, as functions of the
distance of the hole from the center of the nucleus. Note that the probability
for shallow ($E^*<S_n$) holes becomes equal to 1 beyond a certain radius. As
illustrated by Fig.~\ref{fig:rp-correlation}(a), this is due to the fact that
there is a strict minimum kinetic energy for nucleons that are found beyond a
given radius.

Analysis of the INC shows that the impact-parameter distribution of events with
only one INC collision peaks around $7.78$~fm and has a root-mean-square (rms)
deviation of $0.99$~fm. The dotted lines in Fig.~\ref{fig:threshold} are set at
$r=(7.78-0.99)\text{~fm}=6.79\text{~fm}$ and are meant as a guide to the
eye. Roughly speaking, most single-collision reactions take place to the right
of the dotted lines. The same lines are also drawn on
Figs.~\ref{fig:marginal-rp-correlation}(b) and \ref{fig:rp-correlation}.

It is clear from the results displayed in Figure~\ref{fig:threshold} that the
standard \incl\ initial conditions are quite different from the results of the
shell-model calculation: in the surface region, the standard \incl{} probability
to punch a shallow hole in the Fermi sea is sensibly larger than its shell-model
counterparts, which seems to confirm that the excitation energy associated with
the ejection of a proton is underestimated by \incl.

\section{Refinement of the initial conditions}
\label{sec:refin-incl-nucl}

We mentioned in Section~\ref{sec:model-description} that an \incl{} nucleon
moves in a square-well potential whose radius $R(T)$ depends on the nucleon
kinetic energy. The function $R(T)$ is uniquely determined by the choice of the
space density $\rho(r)$ and by the assumption that nucleon momenta are uniformly
distributed in the Fermi sphere. We have shown above that this construction
results in excitation energies for one-collision reactions that are much smaller
than those resulting from the shell model and, arguably, than those suggested by
the available experimental data.

We refine the \incl{} initial conditions by allowing fluctuations in
$R(T)$. We introduce a \emph{fuzziness parameter} $f$ ($0\leq f\leq1$) and a
\emph{fuzzy} square-well radius $R(T;f)$. The precise definition of $R(T;f)$ is
reported in the Appendix, so we limit our exposition to its most important
properties: first, $R(T;f)$ is a random variable. Second, for $f=0$ fluctuations
are suppressed and we recover the standard sharp correlation:
\[
R(T;0)=R(T)\text.
\]
Third, for a given value of $T$, fluctuations in $R(T;f)$ are small if $f$ is
close to zero and they are large if $f$ is close to one. Fourth, the
fluctuations are constructed in such a way that the space density is still given
by $\rho(r)$ and the momentum density is still given by a uniform Fermi
sphere. The construction of the fuzzy \incl{} nucleus is analogous to the
standard preparation algorithm \cite{boudard-incl}. The only difference is that
the radius of the square-well potential is no longer in one-to-one
correspondence with the nucleon energy.

The phase-space structure of the fuzzy initial conditions is illustrated by
Fig.~\ref{fig:rp-correlation}(b), which refers to protons in \lead{} with
fuzziness parameter $f=0.5$. Contrary to Fig.~\ref{fig:rp-correlation}(a), we
see that the density does not drop sharply to zero. Instead, protons of a given
kinetic energy can sometimes be found at much larger distances than in the
standard initial conditions [Fig.~\ref{fig:rp-correlation}(a)]. The
kinetic-energy and space distributions, i.e. the projections of
Fig.~\ref{fig:rp-correlation}(b), are shown in
Fig.~\ref{fig:marginal-rp-correlation} as red lines. By construction, the
kinetic-energy and space distributions are almost indistinguishable from those
of the standard initial conditions. The fluctuations in one of the variables
disappear when integrating over the full domain of the other one. Note however
that different results are obtained if one limits the integration domain to some
sub-range. This is illustrated by the dashed lines in
Fig.~\ref{fig:marginal-rp-correlation}(a), which represent the kinetic-energy
distributions of protons found ``in the surface'', i.e.\ to the right of the
vertical dotted line in Fig.~\ref{fig:marginal-rp-correlation}(b). The
distribution of the standard initial conditions vanishes below a certain energy,
while the fuzzy initial conditions extend much deeper in the Fermi sea.

The fuzzy initial conditions introduce 
additional energy fluctuations for the nucleons found at a given
position. Figure~\ref{fig:threshold} indeed demonstrates that the probability to
punch a shallow surface hole \emph{decreases} for increasing fuzziness, i.e.\
for increasing fluctuations. No value of the fuzziness parameter yields a good
fit to the shell-model result, even if one limits oneself to the surface
region. There is some degree of subjectivity in the choice of the best-fit
values, which are taken to be $f=0.5$ for protons and $f=0.3$ for neutrons. For
\calcium{} (not shown), the best-fit value was taken to be $f=0.3$ for both
protons and neutrons.

Summarizing, we have refined the \incl{} initial conditions in two
respects. First, we have included the possibility of introducing a neutron skin,
as described in Section~\ref{sec:space-densities}. Second, we have introduced
fuzzy initial conditions, which increases energy fluctuations in the nuclear
surface and boosts the probability for deep-nucleon removal in surface
collisions. In the framework of the shell model, this effect is genuinely
quantum-mechanical and it is due to the penetration of the wavefunction in the
classically forbidden region.

\section{Results}\label{sec:results}

\begin{table}
  \caption{Cross sections for one-nucleon removal in \proton-nucleus
    reactions, with the following model variants: (a) standard, (b) standard
    plus skin, (c) standard plus surface fuzziness, (d) standard plus
    skin and surface fuzziness. All values are in mb. Experimental data
    are taken from Refs.~\citenum{chen-ca40,enqvist-lead}. Note that the small
    proton skin in $^{40}$Ca has very little impact on the results
    [(a)${}\simeq{}$(b) and (c)${}\simeq{}$(d)].\label{tab:cross-sections}}
  \centering
  \begin{ruledtabular}
    \begin{tabular}{c|cc|cc}
      & \multicolumn{2}{c|}{565-MeV \proton+\calcium} &
      \multicolumn{2}{c}{1-GeV \proton+\lead}\\
      & $-1p$ & $-1n$ &  $-1p$ & $-1n$ \\
      \hline
      (a) & $54.6$ & $46.4$ & $59.5$ & $82.1$\\
      (b) & $54.6$ & $44.6$ & $50.9$ & $112.0$\\
      (c) & $47.6$ & $40.5$ & $42.1$ & $63.4$\\
      (d) & $47.3$ & $38.2$ & $33.6$ & $83.8$\\
      \hline
      \multirow{2}{*}{exp} & $39.6$ & $34.4$ & $17.6$ & $63.7$\\
      & $\pm7.2$ & $\pm7.7$ & $\pm0.5$ & $\pm9.6$\\
    \end{tabular}
  \end{ruledtabular}
\end{table}

\begin{figure}
  \includegraphics[width=\linewidth]{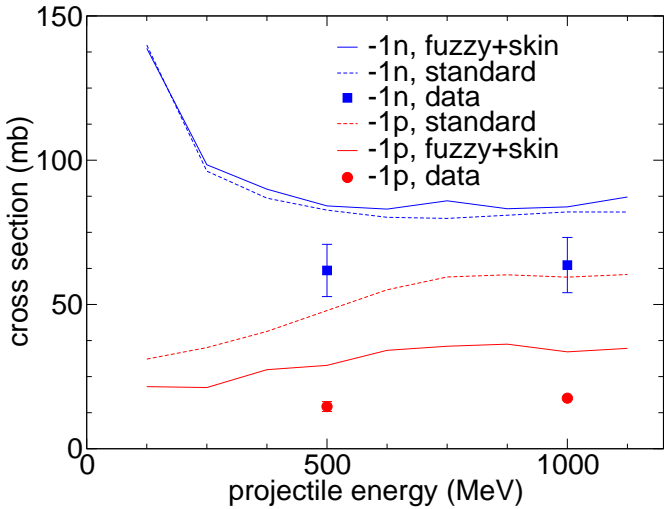}
  \caption{Excitation function for neutron- (blue) and proton-removal (red
    lines) cross sections in \proton+\lead, as calculated with the standard
    \incl{} version (dashed lines) and with the refined treatment of the nuclear
    surface (solid lines). Experimental data are taken from
    Refs.~\citenum{audouin-lead,enqvist-lead}.\label{fig:p_Pb208_excit}}
\end{figure}

\begin{figure}
  \includegraphics[width=\linewidth]{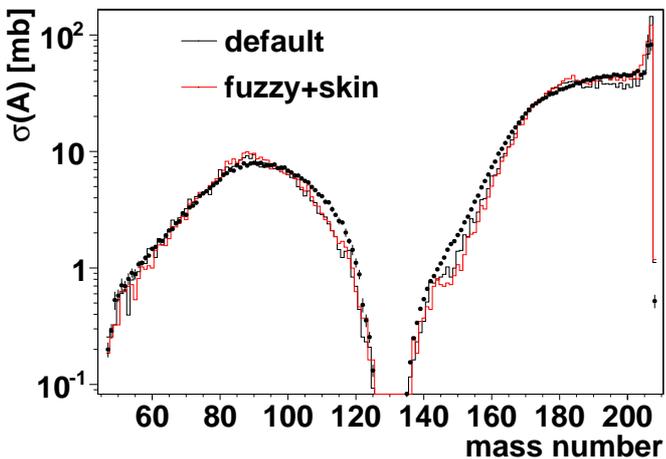}
  \caption{Mass distribution of the residues produced in 1-GeV
    \proton+\lead. The standard \incl{} calculation (black) is compared to the
    refined calculation (red) and to the experimental data
    \cite{enqvist-lead,kelic-bi}.\label{fig:p_Pb208_1000}}
\end{figure}

\begin{figure*}
  \includegraphics[width=\linewidth]{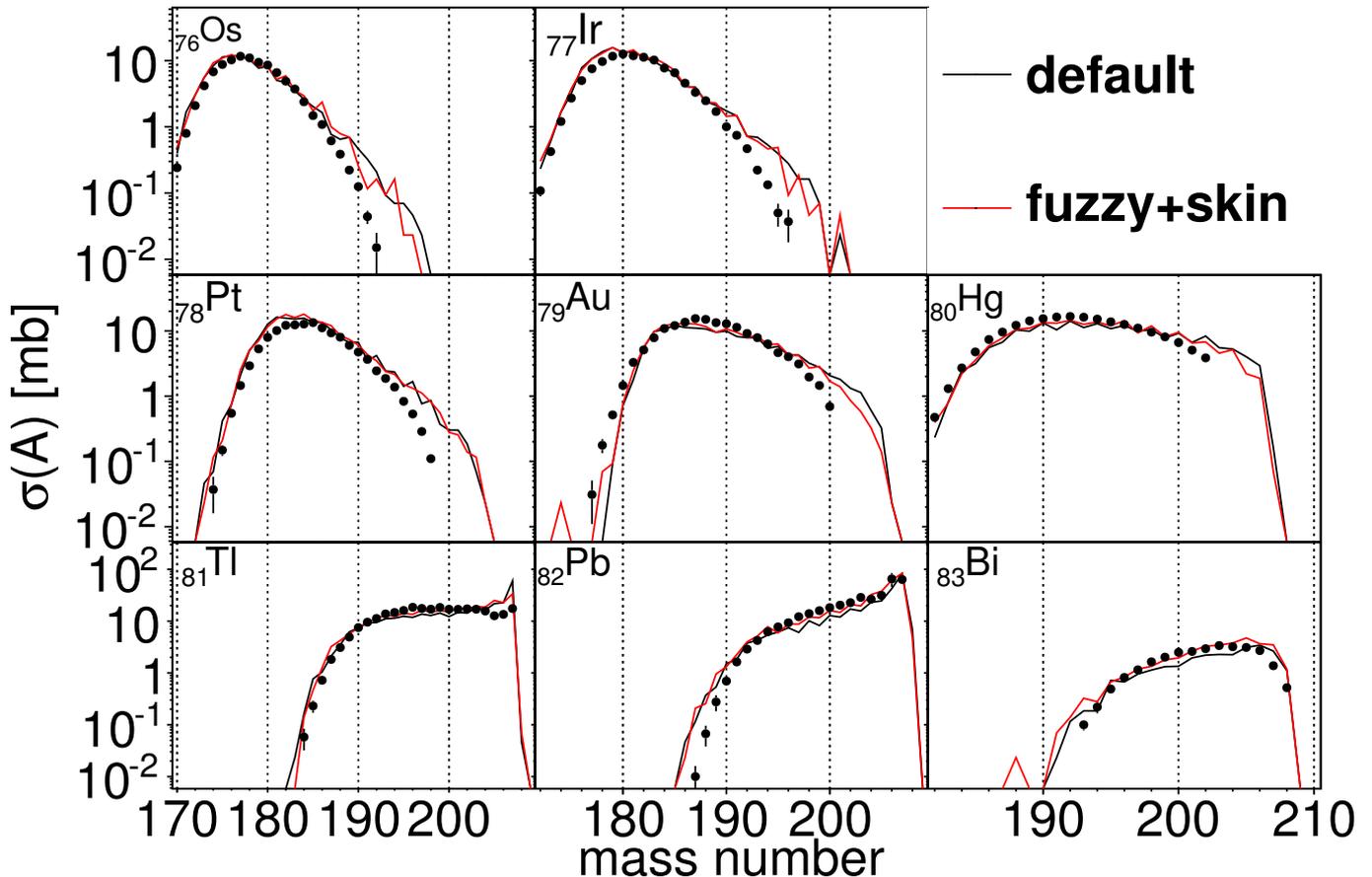}
  \caption{Isotopic distributions of the $Z=76$--$83$ residues
      produced in 1-GeV \proton+\lead. The standard \incl{ calculation (black)
      is compared to the refined calculation (red) and to the experimental data
      \cite{enqvist-lead,kelic-bi}}.\label{fig:p_Pb208_1000_isot_log}}
\end{figure*}

\begin{figure}
  \includegraphics[width=\linewidth]{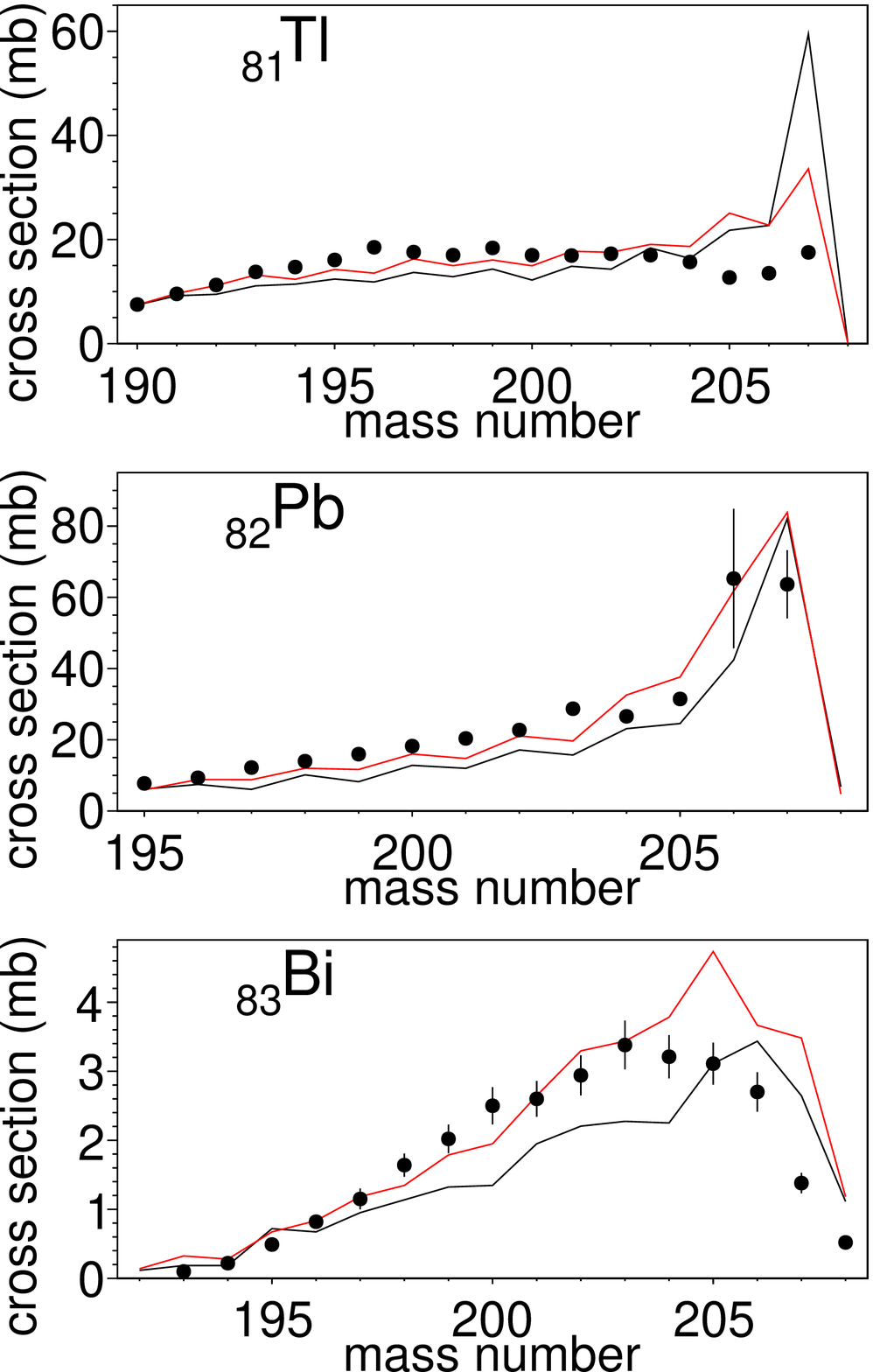}
  \caption{Same as Fig.~\ref{fig:p_Pb208_1000_isot_log}, for $Z=81$--$83$, in
    linear scale.\label{fig:p_Pb208_1000_isot}}
\end{figure}

We turn now to the analysis of the results of the refined INC
model. Table~\ref{tab:cross-sections} shows how the neutron skin and the surface
fuzziness affect the one-nucleon-removal cross sections in \proton+\calcium{}
and \proton+\lead. Unfortunately, no experimental data are available for
\proton+\calcium{} at 1~GeV. There is one experiment at 763~MeV by Chen \etal{}
\cite{chen-ca40}, but the resulting cross sections
($\sigma_{-1p}=(54.7\pm7.9)$~mb; $\sigma_{-1n}=(29.8\pm6.4)$~mb) are in sensible
disagreement with the cross sections measured by the same group at lower
energies and with the values suggested by the systematics of
Fig.~\ref{fig:world-data}. Therefore, we compare our calculations with the
values measured by the same group at the next lowest energy, 565~MeV.

Several observations are due. First, the introduction of the neutron skin in
\lead{} boosts the neutron-removal cross section, as expected. This is however
undesired, since the cross section calculated by standard \incl{} is already in
moderate excess of the experimental value. Second, surface fuzziness suppresses
the cross sections for both channels. This is true both for \calcium{} and
\lead. Third, neither effect is sufficient to compensate for the overestimation
of the proton-removal cross section in \lead\ \emph{if considered alone}.

When the two refinements are simultaneously applied to \lead, the effect of
surface fuzziness for neutron removal almost exactly compensates the effect of
the neutron skin, and the final result ($83.8$~mb) is very close to the value
calculated with standard \incl{} ($82.1$~mb), which is within two standard
deviations (about $30\%$) of the experimental value. The proton-removal cross
section, on the other hand, is reduced by almost a factor of two, which brings
it much closer to the experimental datum, but not quite in agreement with
it.

The results for \calcium{} are qualitatively similar. We observe that the cross
sections are essentially insensitive to the addition of the very thin proton
skin; surface fuzziness, on the other hand, reduces both cross sections by
roughly the same amount (about $15$--$20\%$) and brings them in better agreement
with the experimental data.

The excitation curves for one-nucleon removal are shown in
Fig.~\ref{fig:p_Pb208_excit}. The refined predictions are globally similar to
the standard ones for neutron removal; for proton removal, the excitation
function is roughly rescaled as a whole by a factor of $\sim0.6$. This brings
the prediction in better agreement with the trend shown by the experimental
data.

The global effect of neutron skin and surface fuzziness is partially illustrated
by Fig.~\ref{fig:p_Pb208_1000}, which shows the mass distribution of the
residues produced in 1-GeV \proton+\lead. It is clear that, except for the
$A>170$ region, the refined \incl{} calculation is globally very similar to the
standard result. The fact that the fission peak is essentially unmodified
suggests that neutron skin and surface fuzziness globally do not influence much
the nature of the cascade remnants. Nevertheless, the refined treatment sensibly
ameliorates the cross sections for the heaviest residues ($A>170$), which were
slightly too low in standard \incl.

Insight can be gained by examining the isotopic distributions for the heaviest
residues, which are depicted in Fig.~\ref{fig:p_Pb208_1000_isot_log}. Besides
the cross sections for one-proton ($^{207}$Tl) and one-neutron removal
($^{207}$Pb), the largest differences between the standard and the refined
calculations concern the isotopes of Pb and Bi, which are highlighted (in linear
scale) in Fig.~\ref{fig:p_Pb208_1000_isot}. Lead and bismuth isotopes are
respectively fed by reactions such as $(p,p\,xn)$ and $(p,xn)$, although the
contribution from pionic channels [such as $(p,\pi^+\,xn)^{209-x}\text{Pb}$] is
in general not negligible at all. The refined treatment of the surface
considerably improves the predictions for these cross sections.  Somewhat
surprisingly, the cross sections for the heaviest measured Bi residues
($A=205$--$208$) are degraded. Note however that the production of these
residues imposes constraints on the cascade outcome that are even stricter than
for one-nucleon removal: the excitation energy deposited in the cascade remnant
must be very small, but in addition the incoming proton must be absorbed. This
results in cross sections ($\sim$~mb) which are much smaller than those for
one-nucleon removal ($\sim50$~mb) and which are even more sensitive to the
details of the initial conditions and of the dynamics. Generally, however,
surface fuzziness and neutron skin considerably improve the cross sections for
Pb and Bi isotopes. This should be seen as a by-product of the model refinement
which strengthens our confidence in the new treatment.

\begin{figure}
  \includegraphics[width=\linewidth]{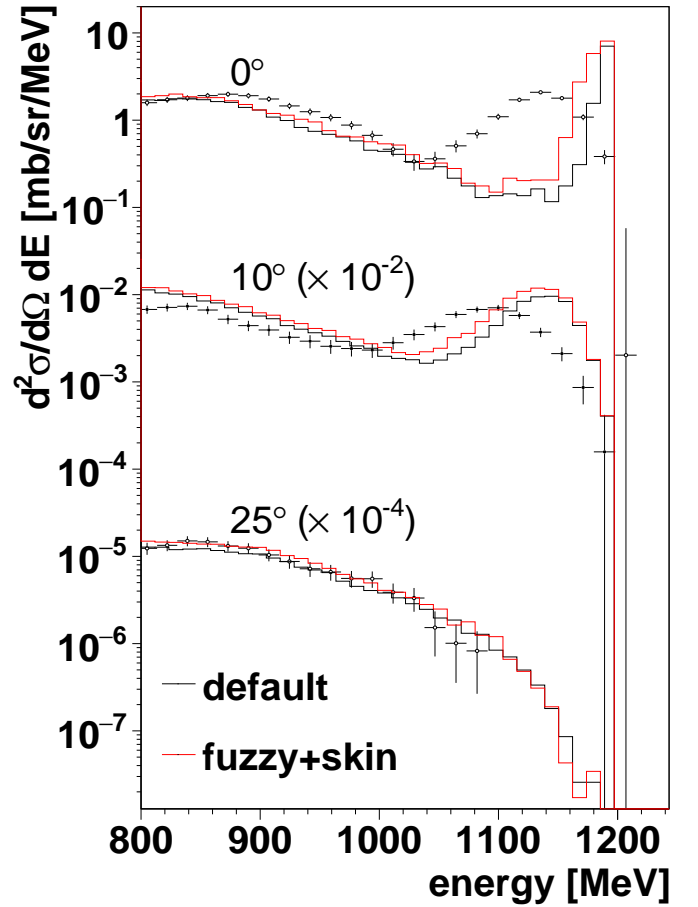}
  \caption{\highlight{Double-differential cross sections for the production of
      neutrons in 1.2-GeV \proton+Pb, as calculated with the standard \incl{}
      version (black lines) and with the refined treatment of the nuclear
      surface (red lines). Experimental data are taken from
      Ref.~\citenum{leray-neutrons}.}\label{fig:p_Pb208_1200_ddxs_n}}
\end{figure}

\highlight{The new description of the surface might also influence other
  observables; for instance, one might expect an effect on the emission patterns
  of particles from peripheral collisions. This is illustrated by the
  double-differential cross sections for neutron production from 1.2-GeV
  \proton+Pb shown in Fig.~\ref{fig:p_Pb208_1200_ddxs_n}. In order to highlight
  the effect of the new surface description, we only show angles below
  $30^\circ$ and energies above $800$~MeV [the effect is much smaller in the
  rest of the neutron momentum space; in general, it is also smaller for
  outgoing protons than for outgoing neutrons (not shown)]. The peak at the
  high-energy end of the $0^\circ$ and $10^\circ$ spectra is due to
  quasi-elastic charge-exchange scattering of the incoming proton off a neutron
  in the target.  The refined treatment of the surface leads to a broadening of
  the quasi-elastic peak, which can easily be understood as a consequence of the
  increased energy fluctuations of the target surface nucleons. However, the
  effect is minor, and it is surely insufficient to reconcile the calculation
  with the experimental data. This well-known disagreement has been known for
  quite some time \cite{boudard-incl} and is probably not specific to the \incl\
  model. Note however that the shape and position of the quasi-elastic peak are
  sensitive at least to the beam profile, the beam energy distribution, the
  target thickness and the detector angular acceptance. None of these aspects is
  realistically modeled in our calculations.}

In summary, we have seen that the proton-removal cross section in 1-GeV
\proton+\lead{} can be reduced by about a factor of two by taking into account
the presence of the neutron skin and the surface fuzziness. However, the refined
value is still in excess of the experimental one by another factor of two. One
might wonder if the results we have obtained can be significantly improved by
refining the calculation of the wavefunctions and of the energy levels (by using
e.g.\ the Hartree-Fock or Hartree-Fock-Bogoliubov methods) \footnote{This will
  anyway be necessary if we wish to generalize our treatment to non-magic
  nuclei.}.  In keeping with the approach described above, we would then need to
fit the refined probability curves with our fuzziness parameter. However, we
have performed a phenomenological exploration of the parameter space and we have
verified that our choice is close to optimal.  It is discouraging to learn that
little can be gained by refining the wavefunctions. Remember however that even
the optimal fuzziness values do not reproduce the shell-model calculations very
well (Fig.~\ref{fig:threshold}). In this sense there is probably margin for
improvement in further refinement of the \incl{} initial conditions, which are
manifestly not flexible enough to fit the shell-model calculations.

We wish to add a short comment about the universality of the failure of INC to
appropriately describe one-nucleon-removal cross sections. The introduction of a
neutron skin can be seen as a straightforward extension of the standard INC
initial conditions; however, we have shown that it is not sufficient to improve
the one-nucleon-removal cross sections. Surface fuzziness, on the other hand,
goes definitely beyond the standard INC initial conditions. The reduction of the
cross sections is ultimately due to the increase of the excitation energy
associated with the knock-out of surface nucleons during the INC
phase. Equivalently, surface nucleons in standard \incl{} are too close to the
Fermi energy to result in deep holes. The larger is the nucleon energy, the
largest is the volume spanned by the trajectory: this assumption is absolutely
natural because it draws from the behavior of classical particles. It is
therefore likely to figure in all INC models where the motion of the individual
target nucleons is explicitly followed.

The discussion above does not apply to intranuclear cascades that model the
Fermi sea as a continuous medium (e.g.\ Isabel). In this case, however,
collisions in low-density regions are dealt with by assuming a reduced value of
the Fermi momentum (so-called \emph{local} Fermi momentum) on top of a reduced
depth of the potential well. Insofar as the depth of the holes that can be
punched in the surface is concerned, the net result is the same: the holes lie
quite close to the Fermi surface and yield rather small excitation energies. In
this sense, the failure of INC can be described as universal, i.e.\ independent
of the specific model incarnation.

Reaction models such as \bertini{} and \cem{} require further discussion. We
have seen that they are also unable to consistently reproduce the experimental
data (Fig.~\ref{fig:world-data}), but the disagreement is qualitatively
different from models such as \incl{} and \isabel. There are prominent
differences between these classes of models: most notably, (i) \bertini{} and
\cem{} are ``space-like'' INCs, while \incl{} and \isabel{} are ``time-like''
INCs; and (ii) \bertini{} and \cem{} include an intermediate pre-equilibrium
stage, whereas \incl{} and \isabel{} are directly coupled to statistical
de-excitation. It is tempting to ascribe the different behavior of the two model
classes to one or both of these elements. We suggest that the use of
pre-equilibrium might be responsible for the different behavior. As we discussed
above, pre-equilibrium models carry no information about the localization of
excitons in configuration space. Of course this is justifiable if the nucleon
wavelength is sufficiently large; nevertheless, the approximation might be too
crude for the description of grazing collisions such as those described in the
present paper.

\section{Conclusions}\label{sec:conclusions}
  
In conclusion, we have shown that INC/de-excitation models universally fail to
describe the cross sections for one-nucleon removal in reactions induced by
high-energy protons. This shortcoming is rather serious, because
INC/de-excitation models are the only viable choice for the description of these
observables. We have used simple shell-model calculations as a guidance for
refining the description of the surface in the INC initial conditions. We
believe we have presented strong arguments indicating that the reason for this
deficiency lies in the presence of neutron skins in heavy, stable nuclei and in
the description of the energy density of the nuclear surface. The refined model,
as it is defined here, introduces no fitting parameters and yields encouraging
predictions: the one-nucleon removal cross sections are substantially improved,
but are still in disagreement with the experimental values for heavy targets. As
a by-product, the isotopic cross sections for the removal of up to several
nucleons are also improved by the refined treatment. Still, further work is
necessary to achieve closer agreement with the experimental data on heavy
nuclei. In the future it will also be necessary to generalize and systematize
our approach to any nucleus, magic or non-magic.

\begin{acknowledgments}
  This work has been partially supported by the EU ENSAR FP7 project (grant
  agreement 262010). We wish to thank Dr.\ Jos\'e Benlliure for stimulating
  discussions and for commenting on the manuscript. We express our gratitude to
  Dr.\ Stepan Mashnik for his enlightening remarks.
\end{acknowledgments}

\appendix*

\section{definition of the fuzzy energy-radius correlation}
\label{sec:defin-fuzzy-energy}

We start by reporting the standard definition of the function that associates
the radius of the square potential well to the nucleon kinetic energy in
\incl. The original equation \cite[][Eq.~(4)]{boudard-incl} is an implicit
definition formulated in terms of the nucleon momentum $p$:
\[ {\left(\frac{p}{p_F}\right)}^3\!\!\!\!\!=
-\frac{4\pi}{3A}\int_0^{\widetilde{R}(p)}\!\!\!\!\!r^3\frac{\ud\rho(r)}{\ud
  r}\ud r\text;
\]
here $p_F$ is the Fermi momentum, $A$ is the mass number of the nucleus and
$\rho(r)$ is the assumed space density. We have slightly modified the notation
in Ref.~\cite{boudard-incl} to read $\widetilde R(p)$. In relation to the
function $R(T)$ that we used in the main text of this paper, it should be
understood that $R(T) = \widetilde{R}\left(\sqrt{T(T+2m)}\right)$. For
conciseness' sake, we omit the indication of the nucleon isospin.

\subsection{Standard algorithm}

The standard \incl{} algorithm for assigning positions and momenta to a nucleon
proceeds as follows:
\begin{enumerate}
\item draw a random momentum $\vec{p}$ from the uniform Fermi sphere; the
  vector direction is isotropic and the absolute value is
  \[
  p=p_F u^{1/3}\text,
  \]
  where $u$ is a uniform random number from the $[0,1]$ interval;
\item compute the associated radius $\widetilde{R}(p)$;
\item draw a random position from a uniform sphere of radius $\widetilde{R}(p)$.
\end{enumerate}
This algorithm trivially results in the following phase-space density [see
Fig.~\ref{fig:rp-correlation}(a)]
\begin{equation}
  \frac{\ud n}{\ud^3\vec{r}\ud^3\vec{p}}=A\frac{\Theta(\widetilde{R}(p)-r)}{(4\pi/3) \widetilde{R}(p)^3}\frac{\Theta(p_F-p)}
  {(4\pi/3) p_F^3}\text,\label{eq:phase-space_standard}
\end{equation}
where $\Theta$ represents the Heaviside step function. It was proven in
Ref.~\citenum{boudard-incl} that Eq.~\eqref{eq:phase-space_standard} has the
appropriate marginal distributions:
\begin{subequations}\label{eq:marginals_standard}
  \begin{align}
    \frac{\ud n}{\ud^3\vec{r}}&{}=\int\!\!\!\frac{\ud n}{\ud^3\vec{r}\ud^3\vec{p}}\ud^3\vec{p}=\rho(r)\text;\\
    \frac{\ud n}{\ud^3\vec{p}}&{}=\int\!\!\!\frac{\ud
      n}{\ud^3\vec{r}\ud^3\vec{p}}\ud^3\vec{r}=A\frac{\Theta(p_F-p)}{(4\pi/3)
      p_F^3}\text.
  \end{align}
\end{subequations}

\subsection{Independent algorithm}

The standard algorithm assigns a unique potential radius to nucleons with a
given momentum. An extreme alternative would be to make the potential radius
completely independent of the momentum. This can be achieved as follows:
\begin{enumerate}
\item draw a random momentum $p$ from the uniform Fermi sphere:
  \[
  p=p_F u^{1/3}\text;
  \]
\item draw another, uncorrelated uniform random number $v$ and define a
  momentum-like variable $p'$:
  \[
  p'=p_F v^{1/3}\text;
  \]
\item use $p'$ to compute the potential radius $\widetilde{R}(p')$;
\item draw a random position from a uniform sphere of radius
  $\widetilde{R}(p')$.
\end{enumerate}
It is easy to prove that the independent algorithm yields the following
phase-space density,
\begin{equation}
\frac{\ud n}{\ud^3\vec{r}\ud^3\vec{p}}=\frac{\rho(r)\Theta(p_F-p)}
{(4\pi/3) p_F^3}\text.\label{eq:phase-space_independent}
\end{equation}
Since Eq.~\eqref{eq:phase-space_independent} factorizes in a space part and a
momentum part, it is trivial to show that it yields the same marginal
distributions as Eq.~\eqref{eq:phase-space_standard}, viz.\
Eqs.~\eqref{eq:marginals_standard}.

\subsection{Fuzzy algorithm}

The two algorithms above can be seen as limiting cases of the following:
\begin{enumerate}
\item draw a random momentum $p$ from the uniform Fermi sphere:
  \begin{subequations}\label{eq:p_pprime}
    \begin{equation}
      p=p_F u^{1/3}\text;
    \end{equation}
  \item draw another \emph{correlated} uniform random number $v$ and define a
    momentum-like variable $p'$:
    \begin{equation}
      p'=p_F v^{1/3}\text;
    \end{equation}
  \end{subequations}
\item use $p'$ to compute the potential radius $\widetilde{R}(p')$;
\item draw a random position from a uniform sphere of radius
  $\widetilde{R}(p')$.
\end{enumerate}
The crucial difference with respect to the independent algorithm is that the
random numbers $u$ and $v$ are \emph{correlated}, i.e.\ they are drawn from
some joint distribution function $g(u,v)$.

The phase-space density generated by the fuzzy algorithm is
\begin{equation}
  \frac{\ud n}{\ud^3\vec{r}\ud^3\vec{p}}=A\frac{\Theta(p_F-p)} {(4\pi/3)p_F^3}
  \int_0^1\!\!\!\ud v\,
  g(u,v)\frac{\Theta(\widetilde{R}(p')-r)}{(4\pi/3)\widetilde{R}(p')^3}\text,
  \label{eq:phase-space_fuzzy}
\end{equation}
where it is understood that $p$ and $p'$ are respectively functions of $u$ and
$v$ through Eqs.~\eqref{eq:p_pprime}. Note that the standard algorithm is
recovered for
\[
g(u,v)=\delta(u-v)\text,
\]
while the independent algorithm results from
\[
g(u,v)=1
\]
(remember that $u,v\in[0,1]$).

The marginal space and momentum distributions can be shown to be
\begin{subequations}\label{eq:marginals_fuzzy}
  \begin{align}
    \frac{\ud n}{\ud^3\vec{r}}&{}=-\int_r^\infty\!\!\ud
    r'\frac{\ud\rho(r')}{\ud r'}\cdot g_{v}\!\!\left({\left(\frac{\widetilde{R}^{-1}(r')}{p_F}\right)}^3\right)\\
    \frac{\ud n}{\ud^3\vec{p}}&{}=A\frac{\Theta(p_F-p)}{(4\pi/3) p_F^3}\cdot
    g_u(u)\text,
  \end{align}
\end{subequations}
where $\widetilde{R}^{-1}$ is the inverse function of $\widetilde{R}$ and $g_u$
and $g_{v}$ are the marginal distributions of $g$:
\begin{align*}
  g_u(u)&{}=\int_0^1\!\!\ud v\,g(u,v)\\
  g_{v}(v)&{}=\int_0^1\!\!\ud u\,g(u,v)\text.
\end{align*}
Equations~\eqref{eq:marginals_fuzzy} demonstrate that the fuzzy algorithm
generates the appropriate marginal space and momentum distributions
(Eqs.~\eqref{eq:marginals_standard}) if and only if the marginals of $g$ are
uniform:
\begin{align*}
  g_u(u)&{}=1\\
  g_{v}(v)&{}=1\text.
\end{align*}

\subsubsection{Construction of the joint distribution $g(u,v)$}

Having characterized the conditions for recovering the correct space and
momentum distributions, we now show how to construct a joint distribution on the
unit square with uniform marginals. We would like the $u$-$v$ correlation to be
continuously ``tunable'' between the two extreme cases of the standard and the
independent algorithm. We therefore introduce a \emph{fuzziness parameter} $f$
and denote the joint distribution as $g(u,v;f)$.

There are several solutions to this deceptively simple problem. The one we have
adopted in \incl, in a nutshell, consists in generating two correlated normal
deviates (which can be done with a simple algorithm) and mapping them to the
unit square using the inverse of the normal cumulative distribution
function. The method is a simple application of the theory of copulas
\cite{nelsen-copulas}.

In detail, we start out with a bivariate standard normal distribution with
correlation coefficient $c$, which can explicitly be written as
\begin{gather*}
  h(w,z;c)=\frac{1}{2\pi\sqrt{1-c^2}}\,\eta(w,z;c)\\
  \eta(w,z;c)=\exp{\left[-\frac{w^2+z^2-2cwz}{2(1-c^2)}\right]}\text.
\end{gather*}
Both $w$ and $z$ are standard normal variables:
\begin{subequations}\label{eq:standard_normal_marginals}
  \begin{align}
    \int h(w,z;c)\ud z&{}=\frac{1}{\sqrt{2\pi}}\exp(-w^2/2)\\
    \int h(w,z;c)\ud w&{}= \frac{1}{\sqrt{2\pi}}\exp(-z^2/2)\text.
  \end{align}
\end{subequations}
It is easy to show (by factorization) that sampling from $h(w,z;c)$ can be
performed as follows: first sample $w$ from a standard normal distribution, then
sample $z$ from a normal distribution with mean $cw$ and variance $1-c^2$. 
We finally define
\begin{align*}
  u&{}=\Phi(w)\\
  v&{}=\Phi(z)\text,
\end{align*}
where $\Phi(x)$ is the cumulative distribution function of the standard normal
distribution. This maps the standard normal random variables $(w,z)$ onto the
(open) unit square mapped by $(u,v)$. Since $w$ and $z$ are standard normal
variables (Eqs.~\eqref{eq:standard_normal_marginals}), $u$ and $v$ are uniform
over the unit interval.

The variables $w$ and $z$ are not independent; therefore, neither are $u$ and
$v$. However, we were unable to derive a closed expression for the correlation
coefficient of $g(u,v;f)$. We identify the fuzziness parameter $f$ with $1-c$,
so that for $c=0$ (uncorrelated variables) we have $f=1$ (full fuzziness) and
for $c=1$ (perfect correlation) we have $f=0$ (no fuzziness). The joint
distribution can be shown to be
\[
g(u,v;f)={\left[\sqrt{f(2-f)}\,\eta\!\left(\Phi^{-1}(u),\Phi^{-1}(v);\frac{1}{1-f}\right)\right]}^{-1}\text.
\]

The construction can be straightforwardly generalized to any other initial joint
distribution $h(w,z)$ for which a simple sampling algorithm exists.

\subsubsection{Definition of the fuzzy radius}

\begin{figure}
  \centering
  \includegraphics[width=\linewidth]{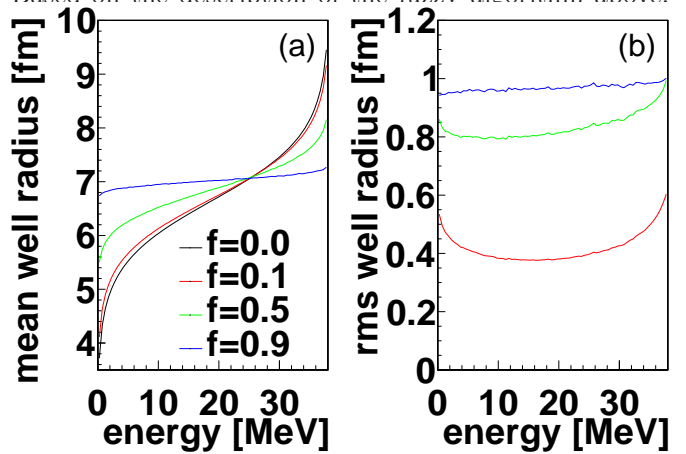}
  \caption{Mean (a) and root-mean-square deviation (b) of the probability
    distribution for the fuzzy well radius, as functions of the nucleon kinetic
    energy, for different values of the fuzziness parameter $f$. The rms
    deviation vanishes for $f=0$ (standard
    algorithm).\label{fig:fuzziness-mean-rms}}
\end{figure}

We conclude by reporting the explicit definition for the function $R(T;f)$ that
we used in the text:
\[
R(T;f)=\widetilde{R}\left(\sqrt{T(T+2m)};f\right)\text.
\]
Here $\widetilde{R}(p;f)$ must be understood as a random variable. Based on the
description of the fuzzy algorithm above, the probability that
$\widetilde{R}(p;f)$ assumes the value $\xi$, for a given momentum $p$, can
be written as
\begin{multline}\label{eq:pdf_fuzzy_radius}
  \frac{\ud P(\xi<\widetilde{R}(p;f)\leq \xi+\ud\xi)}{\ud\xi}= g(u,v;f)\frac{\ud
    v}{\ud
    p'}\frac{\ud p'}{\ud \widetilde{R}(p')}\\
  {}=\frac{3p'^2}{\widetilde{R}'(p')\,p_F^3}
  g\!\left({\left(p/p_F\right)}^3\!,{\left(p'/p_F\right)}^3;f\right)\text,
\end{multline}
where it should be understood that $p'=\widetilde{R}^{-1}(\xi;f)$.

As an illustration, Fig.~\ref{fig:fuzziness-mean-rms} shows the mean and
root-mean-square deviation of the probability distribution given by
Eq.~\eqref{eq:pdf_fuzzy_radius}, for different values of the fuzziness parameter
$f$. It is worth stressing that not only the rms but also the mean values depend
on $f$; it is clear that it must be so because for $f=1$ we must recover the
independent algorithm, in which the well radius is independent of the nucleon
energy.

\end{document}